\documentclass[twocolumn,prl,superscriptaddress]{revtex4}
\usepackage{amsfonts}
\usepackage{amsmath}
\usepackage{amssymb}
\usepackage{graphicx}
\usepackage{bm}
\usepackage{color}
\usepackage{mathrsfs}
\usepackage{amsbsy}
\usepackage{latexsym,epsfig,graphicx}
\usepackage{dcolumn}
\usepackage{subfigure}
\usepackage{comment}
\usepackage{xspace}
\usepackage{epstopdf}
\usepackage{tabularx}
\usepackage{longtable}
\usepackage[colorlinks=true, letterpaper=true, pdfstartview=FitV, linkcolor=blue, citecolor=blue, urlcolor=blue]{hyperref}
\usepackage[normalem]{ulem}
\newcommand{\RNum}[1]{\uppercase\expandafter{\romannumeral #1\relax}}

\begin{document}
\title{Exact solutions of non-Hermitian  chains with asymmetric long-range hopping under specific boundary conditions}

\author{Cui-Xian Guo}
\affiliation{Beijing National Laboratory for Condensed Matter Physics,
Institute of Physics, Chinese Academy of Sciences, Beijing 100190, China}


\author{Shu Chen}
\email{schen@iphy.ac.cn}
\affiliation{Beijing National Laboratory for Condensed Matter Physics,
Institute of Physics, Chinese Academy of Sciences, Beijing 100190, China}
\affiliation{School of Physical Sciences, University of Chinese Academy of Sciences, Beijing 100049, China}
\affiliation{Yangtze River Delta Physics Research Center, Liyang, Jiangsu 213300, China}

\begin{abstract}
We study one-dimensional general non-Hermitian models with asymmetric long-range hopping and explore to analytically solve the systems under some specific boundary conditions. Although the introduction of long-range hopping terms prevents us from finding analytical solutions for arbitrary boundary parameters, we identify the existence of exact solutions when the boundary parameters fulfill some constraint relations, which give the specific boundary conditions.  Our analytical results show that the wave functions take simple forms and are independent of hopping range, while the eigenvalue spectra display rich model-dependent structures. Particularly, we find the existence of a special point coined as pseudo-periodic boundary condition, for which the eigenvalues are the same as the periodical system when the hopping parameters fulfill certain conditions, whereas eigenstates display non-Hermitian skin effect.
\end{abstract}
\maketitle

%
%

\section{Introduction}
Recently, non-Hermitian systems have gained much attention, both theoretically and experimentally \cite{Ueda,RMP}.
In contrast to Hermitian systems, non-Hermitian systems exhibit many novel properties, such as complex spectrum structures, rich topological classifications and non-Hermitian skin effect (NHSE) \cite{Sato,Zhou,Gong,CHLiu1,CHLiu2,SYao2,Kunst,KZhang,KYokomizo,LeeCH,Okuma,HShen,Yin,Xiong,Alvarez,TELee,Leykam,SYao1,Jiang2018}. NHSE is characterized by the emergence of a large number of bulk states accumulating on one of the open boundaries accompanying with remarkably different spectra from those under periodic boundary condition (PBC)  \cite{SYao2,Kunst,KZhang,KYokomizo,LeeCH,Okuma,HShen}. As this counterintuitive phenomenon has no Hermitian correspondence,  the NHSE has attracted intensive studies in the past years
\cite{Slager,HJiang,WYi,LJin,Kou,Longhi-PRR,Herviou,ZSYang,GongJB,Ezawa,Imura,LeeCH-PRB2020}.

The NHSE is essentially a boundary-sensitivity phenomenon. The boundary effect for non-Hermitian systems has been studied in Ref.\cite{Slager,LeeCH,Roccati,GuoCX,LiuYX-2021,YXLiu,Jiang-EPJB,Turker,Budich-EPJD,LangLJ,Linhu}. To understand why the change of boundary terms dramatically affects the properties of bulk states of non-Hermitian systems, with collaborators we presented exact solutions for one-dimensional non-Hermitian models with generalized boundary conditions in a recent work \cite{GuoCX}, in which size-dependent boundary effect has been clarified from the perspective of exact solution.
The analytical results uncovered the existence of size-dependent NHSE and gave quantitative description of the interplay effect of boundary hopping terms and lattice size.
The size-dependent NHSE was firstly observed by Li et al. in coupled nonreciprocal chains \cite{CSE},  and was generalized to open quantum systems \cite{CHLiu2020}.  It was also observed in non-reciprocal chains with impurity \cite{LiuYX-2021}.

In this paper, we generalize the exact solutions to one-dimensional non-Hermitian models with nonreciprocal (asymmetric) long-range hopping under specific boundary conditions.
Although the introduction of long-distance hopping terms hinders the finding of analytical solutions for arbitrary boundary parameters, we identify the existence of exact solutions when the boundary parameters fulfill some constraint relations, which give the specific boundary conditions considered in the present work. Under the specific boundary conditions, we exactly solve the eigenvalue equations and give analytical results of eigenvalues and wavefunctions. Based on our analytical results, we demonstrate the existence of size-dependent NHSE and rich structures of eigenvalue spectra. Some concrete examples are also discussed.

\section{Models and solutions}
\begin{figure}[b]
\begin{center}
\includegraphics[width=0.45\textwidth]{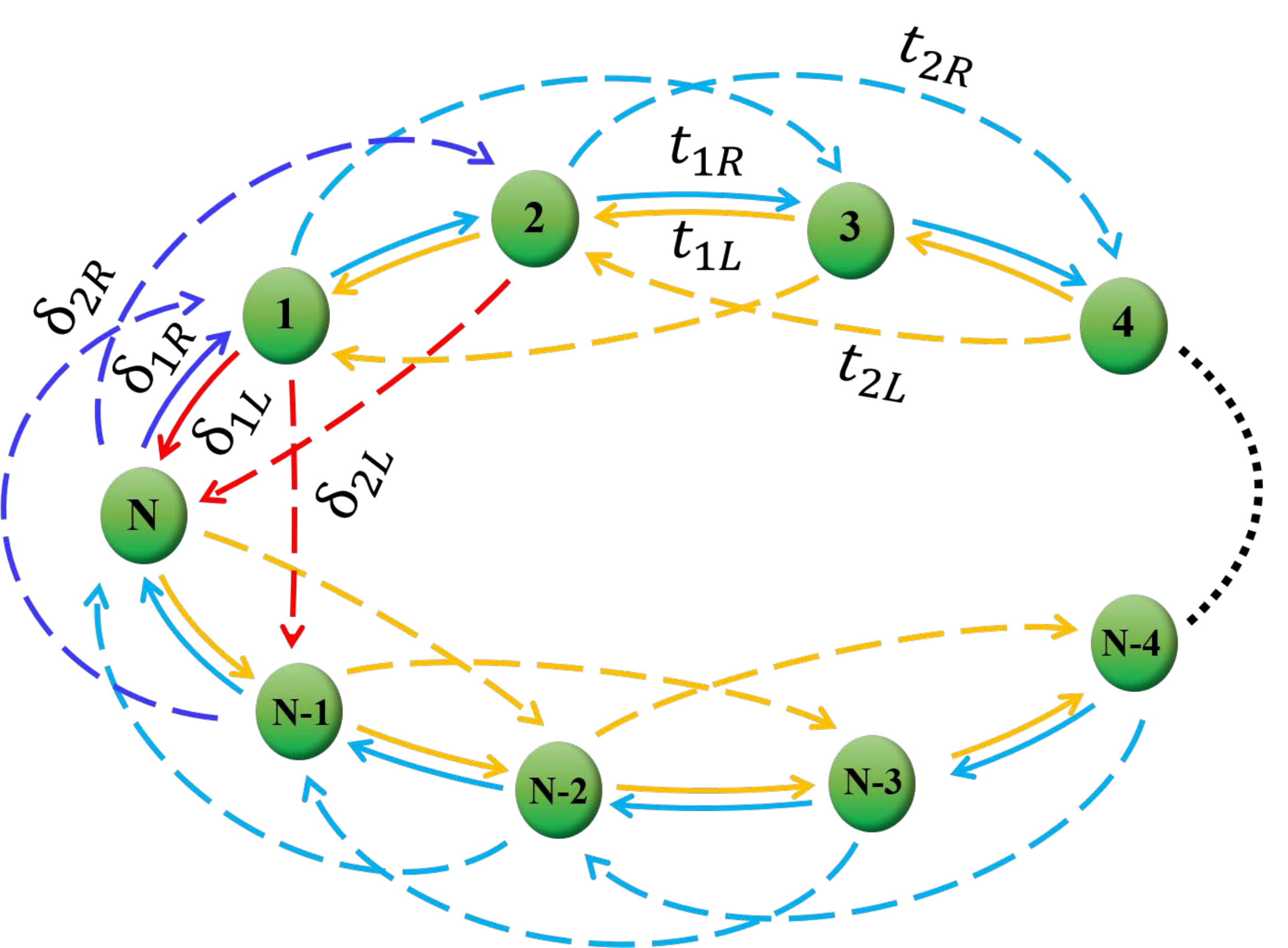}
\caption{Schematic diagram of general 1D non-Hermitian model with $p=2$ and $q=2$.}%
\end{center}
\label{fig1}
\end{figure}

We start with the general 1D non-Hermitian model with asymmetric long-distance hopping terms under generalized boundary conditions, described by the Hamiltonian as follows
\begin{equation}\label{GH}
\begin{split}
\hat{H}=&\sum\limits_{n=1}^{N-j} \left\{ \sum\limits_{j=1}^{p}\left[ t_{jL}\hat{c}_{n}^{\dag
}\hat{c}_{n+j} \right]+ \sum\limits_{j=1}^{q} \left[ t_{jR}\hat{c}_{n+j}^{\dag }\hat{c}_{n} \right] \right\}\\
&+\sum\limits_{n=1}^{j} \left\{ \sum\limits_{j=1}^{p} \left[ \delta_{jL}\hat{c}_{N+n-j}^{\dag }\hat{c}_{n} \right] + \sum\limits_{j=1}^{q}  \left[ \delta_{jR}\hat{c}_{n}^{\dag }\hat{c}_{N+n-j} \right] \right\},
\end{split}
\end{equation}
where $p$ is the farthest length of left hopping, $q$ is the farthest length of right hopping, and $N$ is the number of lattice sites. A model with $p=2$ and $q=2$ is schematically  displayed in Fig. 1.   While the PBC corresponds to $\delta_{jL}=t_{jL} ~(j=1,...,p)$ and $\delta_{jR}=t_{jR} ~ (j=1,...,q)$, the open boundary condition (OBC) corresponds to $\delta_{jL}=\delta_{jR}=0$.

For the system under the PBC, we can perform the following Fourier transformation
\begin{equation}
\hat{c}_{n} = \frac{1}{\sqrt{N}}\sum_{k}  e^{i k n} \hat{c}_{k},~~~~
\hat{c}_{n}^{\dag} = \frac{1}{\sqrt{N}}\sum_{k} e^{-i k n} \hat{c}_{k}^{\dag}.
\end{equation}
Then the Hamiltonian becomes $\hat{H}=\sum_{k}\hat{H}(k)$, where
\begin{equation}
\hat{H}(k)= \sum\limits_{j=1}^{p}  \left[ t_{jL}e^{ikj}\hat{c}_{k}^{\dag
}\hat{c}_{k} \right]+
\sum\limits_{j=1}^{q}\left[ t_{jR}e^{-ikj}\hat{c}_{k}^{\dag }\hat{c}_{k} \right]
\end{equation}
with $k=\frac{2m\pi}{N}~(m=1,2...,N)$.
Thus, the eigenvalue of the general model under PBC is given by
\begin{equation}\label{GEPBC}
E(k)=\sum\limits_{j=1}^{p}  t_{jL}e^{ikj} +\sum\limits_{j=1}^{q}t_{jR}e^{-ikj} .
\end{equation}
For the nonreciprocal lattice, we have $t_{jL} \neq t_{jR}$, and thus the periodic boundary spectrum is complex.

For the general case with $\delta_{jL} \neq t_{jL}$ and $\delta_{jR} \neq t_{jR}$, we need solve the eigenvalue equation for Eq.(\ref{GH}) in real space, which consists of a series of bulk equations and boundary equations. The bulk equations can be expressed as
\begin{equation}\label{GBQ}
\sum\limits_{j=0}^{q-1}\left[ t_{(q-j)R}\psi _{s+j} \right] -E\psi _{s+q}+ \sum\limits_{j=1}^{p}\left[ t_{jL}\psi _{s+q+j} \right] =0
\end{equation}
with $s=1,2,\cdots ,N-(q+p)$. The boundary equations can be expressed as
\begin{equation}\label{GEQq1}
\begin{split}
&\sum\limits_{j=1}^{s-1}\left[ t_{(s-j)R}\psi _{j} \right] -E\psi _{s}+ \sum\limits_{j=1}^{p}\left[ t_{jL}\psi _{s+j} \right]\\
&+ \sum\limits_{j=0}^{q-s}\left[ \delta_{(q-j)R}\psi _{N+s-(q-j)} \right]=0
\end{split}
\end{equation}
with $s=1,2,\cdots ,q$, and
\begin{equation}\label{GEQq2}
\begin{split}
&\sum\limits_{j=1}^{s}\left[ \delta_{(p-s+j)L}\psi _{j} \right]+ \sum\limits_{j=0}^{q-1}\left[ t_{(q-j)R}\psi _{N-p+s-(q-j)} \right] \\
& -E\psi _{N-p+s}+ \sum\limits_{j=1}^{p-s}\left[ t_{jL}\psi _{N-p+s+j} \right]  =0
\end{split}
\end{equation}
with $s=1, 2, \cdots ,p$.

By comparing Eq.(\ref{GEQq1}) with $s=1,2,\cdots ,q$ and Eq.(\ref{GBQ}) with $s=-q+1,-(q-1)+1,\cdots ,0$ respectively, we find that Eq.(\ref{GEQq1}) are equivalent to boundary equations as follows
\begin{equation}\label{GEQ1}
\sum\limits_{j=0}^{q-s}\left[ t_{(q-j)R}\psi _{s-(q-j)} \right]=\sum\limits_{j=0}^{q-s}\left[ \delta_{(q-j)R}\psi _{N+s-(q-j)} \right]
\end{equation}
with $s=1,2,\cdots ,q$. Similarly, by comparing Eq.(\ref{GEQq2}) with $s=1,2,\cdots ,q$ and Eq.(\ref{GBQ}) with $s=N-(q+p)+1,N-(q+p)+2,\cdots ,N-q$ respectively, we find that Eq.(\ref{GEQq2}) are equivalent to boundary equations as follows
\begin{equation}\label{GEQ2}
\sum\limits_{j=1}^{s}\left[ \delta_{(p-s+j)L}\psi _{j} \right]  =   \sum\limits_{j=1}^{s}\left[ t_{(p-s+j)L}\psi _{N+j} \right]
\end{equation}
with $s=1, 2, \cdots ,p$. It is noticed that Eq.(\ref{GBQ}) with $s=-q+1,-(q-1)+1,\cdots ,0$ and $s=N-(q+p)+1,N-(q+p)+2,\cdots ,N-q$ can be viewed as a continuation of bulk equations, therefore the resulting wawefunctions of $\psi_{j}$ with $j=-q+1,-(q-1)+1,\cdots ,0$ and $j=N+1,N+2,\cdots ,N+p$ are auxiliary wavefunctions satisfied Eq.(\ref{GBQ}), just for the purpose of simplifying the calculation.


Due to spatial translational property from bulk equations, we can set the ansatz of wave function $\Psi _{i}$ which satisfies the bulk equations as follows
\begin{equation}\label{GFi}
\Psi _{i} =(z_{i},z_{i}^{2},z_{i}^{3},\cdots ,z_{i}^{N-1},z_{i}^{N})^{T} .
\end{equation}
By inserting Eq.(\ref{GFi}) into the bulk equations Eq.(\ref{GBQ}), we obtain the expression of eigenvalue in terms of $z_i$:
\begin{equation}\label{GEZ}
E=\sum\limits_{j=1}^{p}  t_{jL}z_i^{j} +\sum\limits_{j=1}^{q}t_{jR}z_i^{-j}
\end{equation}
For a given $E$, there are $q+p$ solutions $z_i$ ($z_1, z_2, \cdots, q+p$). Then it follows that the superposition of $p+q$ linearly independent solutions is also the solution of Eq.(\ref{GBQ}) corresponding to the same eigenvalue, i.e.,
\begin{equation}
\Psi=c_{1}\Psi _{1}+c_{2}\Psi _{2}+\cdots+c_{q+p}\Psi _{q+p} = (\psi _{1},\psi _{2},\cdots ,\psi _{N})^{T}~~~ \label{GWave}
\end{equation}
where
\begin{equation}
\psi _{n}=\sum_{i=1}^{q+p}(c_{i}z_{i}^{n})=c_{1}z_{1}^{n}+c_{2}z_{2}^{n}+\cdots+c_{q+p}z_{q+p}^{n}
\end{equation}
with $n=1,2,\cdots ,N$.

To solve the eigenequation $H \Psi = E \Psi$, the general ansatz of wave function should also fulfill the boundary conditions.
By inserting  the expression of $\Psi$ into Eqs.(\ref{GEQ1},\ref{GEQ2}),  the boundary equations can be represented as
\begin{equation}\label{GBQM}
H_{B} (c_{1}, \cdots, c_{q}, c_{q+1}, \cdots, c_{q+p})^{T}=0 .
\end{equation}
Here $H_{B}$ is the boundary matrix given by
\begin{equation*}
\left(
\begin{array}{cccccc}
F_1(z_1) & \cdots   &  F_1(z_q)  &  F_1(z_{q+1})   & \cdots  & F_1(z_{q+p})\\
\vdots   & \vdots   &  \vdots    & \vdots      & \vdots  & \vdots   \\
F_q(z_1) & \cdots   &  F_q(z_q)  &  F_q(z_{q+1})   & \cdots  & F_q(z_{q+p})\\
F_{q+1}(z_1) & \cdots   &  F_{q+1}(z_q)  &  F_{q+1}(z_{q+1})   & \cdots  & F_{q+1}(z_{q+p})\\
\vdots   & \vdots   &  \vdots    & \vdots      & \vdots  & \vdots   \\
F_{q+p}(z_1) & \cdots   &  F_{q+p}(z_q)  &  F_{q+p}(z_{q+1})   & \cdots  & F_{q+p}(z_{q+p})
\end{array}%
\right),
\end{equation*}
where
$$F_s(z_i)=\sum\limits_{j=0}^{q-s}\left[ t_{(q-j)R}-\delta_{(q-j)R}z_i^{N} \right] z_i^{s-(q-j)}$$  with $s=1,\cdots ,q$ and
$$F_{q+s}(z_i)=\sum\limits_{j=1}^{s}\left[ \delta_{(p-s+j)L}  -  t_{(p-s+j)L}z_i^{N} \right]z_i^j$$  with $s=1,\cdots ,p$
and $i=1,\cdots,q+p$. In the above calculation, the auxiliary wavefunctions defined as $\psi _{j}=\sum_{i=1}^{q+p}(c_{i}z_{i}^{j})=c_{1}z_{1}^{j}+c_{2}z_{2}^{j}+\cdots+c_{q+p}z_{q+p}^{j}$ with $j=-q+1,-(q-1)+1,\cdots ,0$ and $j=N+1,N+2,\cdots ,N+p$ are used. Alternatively, the Eq.(\ref{GBQM}) can also be obtained by inserting  the expression of $\Psi$ into Eqs.(\ref{GEQq1},\ref{GEQq2}) in combination with Eq.(\ref{GEZ}).
The nontrivial solutions for ($c_1, c_2, \cdots, c_{q+p}$) mean that $c_1 = 0, c_2 = 0, \cdots, c_{q+p}=0$ cannot be satisfied simultaneously.  The condition for the existence of nontrivial solutions for $(c_1, c_2, \cdots, c_{q+p})$ is determined by
\begin{equation}
\mathrm{det}[H_{B}] =0 ,
\end{equation}
which is usually too complicated to be precisely solved for the general case. For convenience, we shall divide the solutions into two cases: one is that the number of $c_i\neq0 ~ (i=1,...,q+p)$ is 1, and the other is that the number of $c_i\neq0$ is greater than 1 and less than or equal to $q+p$. In general, the second case is hard to be analytically solved. An exception is the case of $p=q=1$, which was exactly solved for arbitrary boundary parameters \cite{GuoCX}.

The solutions of $z_i$ for the first case (i.e. there is only one nonzero $c_i$ and for convenience we denote it as $c_1$) can be easily obtained by applying a simplified method, which is the situation studied in this paper. In this case, eigenfunction is composed of only one solution, i.e., $|\Psi\rangle= c_1|\Psi_1\rangle$, and the boundary
equation $H_B(c_1,\cdots,0)^T = 0$ requires $c_1\neq0, c_2=0, \cdots, c_{q+p}=0$. Thus Eq.(\ref{GBQM}) gives rise to
\begin{equation}
F_1(z_1)=\cdots=F_q(z_1)=F_{q+1}(z_1)=\cdots=F_{q+p}(z_1)=0
\end{equation}
i.e., the following equations should be satisfied simultaneously:
\begin{eqnarray} \label{GQ1BQ1}
\sum\limits_{j=0}^{q-s}\left[ t_{(q-j)R}-\delta_{(q-j)R}z_1^{N} \right] z_1^{s-(q-j)}=0
\end{eqnarray}
with $s=1,2,\cdots ,q$ and
\begin{eqnarray} \label{GQ1BQ2}
\sum\limits_{j=1}^{s}\left[ \delta_{(p-s+j)L}  -  t_{(p-s+j)L}z_1^{N} \right]z_1^j =0
\end{eqnarray}
with $s=1,2,\cdots ,p$.

From Eq.(\ref{GQ1BQ1}) with $s=q$, we can obtain the solution of $z_1$ as
\begin{equation}\label{GQ1Z}
z_1=\sqrt[N]{\mu }e^{i\frac{2m\pi }{N}},
\end{equation}
with
\begin{equation}\label{GQ1BQ2R}
\mu = \frac{t_{qR}}{\delta _{qR}}
\end{equation}
and $m=1,\cdots ,N$.
Then we insert Eqs.(\ref{GQ1Z}) and (\ref{GQ1BQ2R}) into Eq.(\ref{GQ1BQ1}) with $s=q-1,q-2,...,1$ successively, and we have
\begin{equation}\label{GQ1BQ1R}
\frac{t_{(q-1)R}}{\delta _{(q-1)R}}=\cdots=\frac{t_{2R}}{\delta _{2R}}=\frac{t_{1R}}{\delta _{1R}}=\mu .
\end{equation}
Next, we insert Eq.(\ref{GQ1Z}) into Eq.(\ref{GQ1BQ2}) with $s=1$, and we get
\begin{equation}\label{GQ1BQ2L}
\frac{\delta _{pL}}{t_{pL}}=\mu .
\end{equation}
Then we insert Eqs.(\ref{GQ1Z}) and (\ref{GQ1BQ2L}) into Eq.(\ref{GQ1BQ2}) with $s=1,2,...p-1$ successively, and we have
\begin{equation}\label{GQ1LPL}
\frac{\delta _{(p-1)L}}{t_{(p-1)L}}=\cdots=\frac{\delta _{2L}}{t_{2L}}=\frac{\delta _{1L}}{t_{1L}}=\mu .
\end{equation}
Combining Eqs.(\ref{GQ1BQ2R} - \ref{GQ1LPL}) together, we have
\begin{equation}\label{Gcdt}
\frac{t_{1R}}{\delta _{1R}}\cdots=\frac{t_{qR}}{\delta _{qR}}=\frac{\delta _{1L}}{t_{1L}}=\cdots=\frac{\delta _{pL}}{t_{pL}}=\mu,
\end{equation}
which is the specific boundary condition corresponding to the first case. As long as  the specific boundary condition Eq.(\ref{Gcdt}) is fulfilled, the solution of $z_1$ is given by Eq.(\ref{GQ1Z}). We note that $\mu=1$ corresponds to the periodic boundary condition. While we have always $|z_1|=1$ under the PBC, for the general case with $\mu\neq1$, $|z_1|=\sqrt[N]{\mu}$ is not equal to 1.

By inserting Eq.(\ref{GQ1Z}) into Eqs.(\ref{GEZ},\ref{GWave}), we obtain the enigenvalues and eigenfunctions for the general model under the specific boundary condition as
\begin{equation}\label{GEZs}
E=\sum\limits_{j=1}^{p}  t_{jL}(\sqrt[N]{\mu}e^{i\theta})^{j} +\sum\limits_{j=1}^{q}t_{jR}(\sqrt[N]{\mu}e^{i\theta})^{-j}
\end{equation}
and
\begin{equation}\label{GFFis}
\Psi= \left(\sqrt[N]{\mu }e^{i\theta},\left( \sqrt[N]{\mu }e^{i\theta} \right)^{2},\cdots,\left( \sqrt[N]{\mu }e^{i\theta} \right)^{N} \right)^{T},
\end{equation}
where $\theta=\frac{2m\pi}{N}$. When $\mu=1$,  Eq.(\ref{GEZs}) is identical to Eq.(\ref{GEPBC}), and the eigenstates are all extended states, corresponding to PBC.
The case of $\mu= e^{i \phi}$ with $\phi \in (0, 2 \pi)$ corresponds to a twist boundary condition by shifting the momentum a twist angle $\phi/N$.

\begin{figure}[tbp]
\begin{center}
\includegraphics[width=0.47\textwidth]{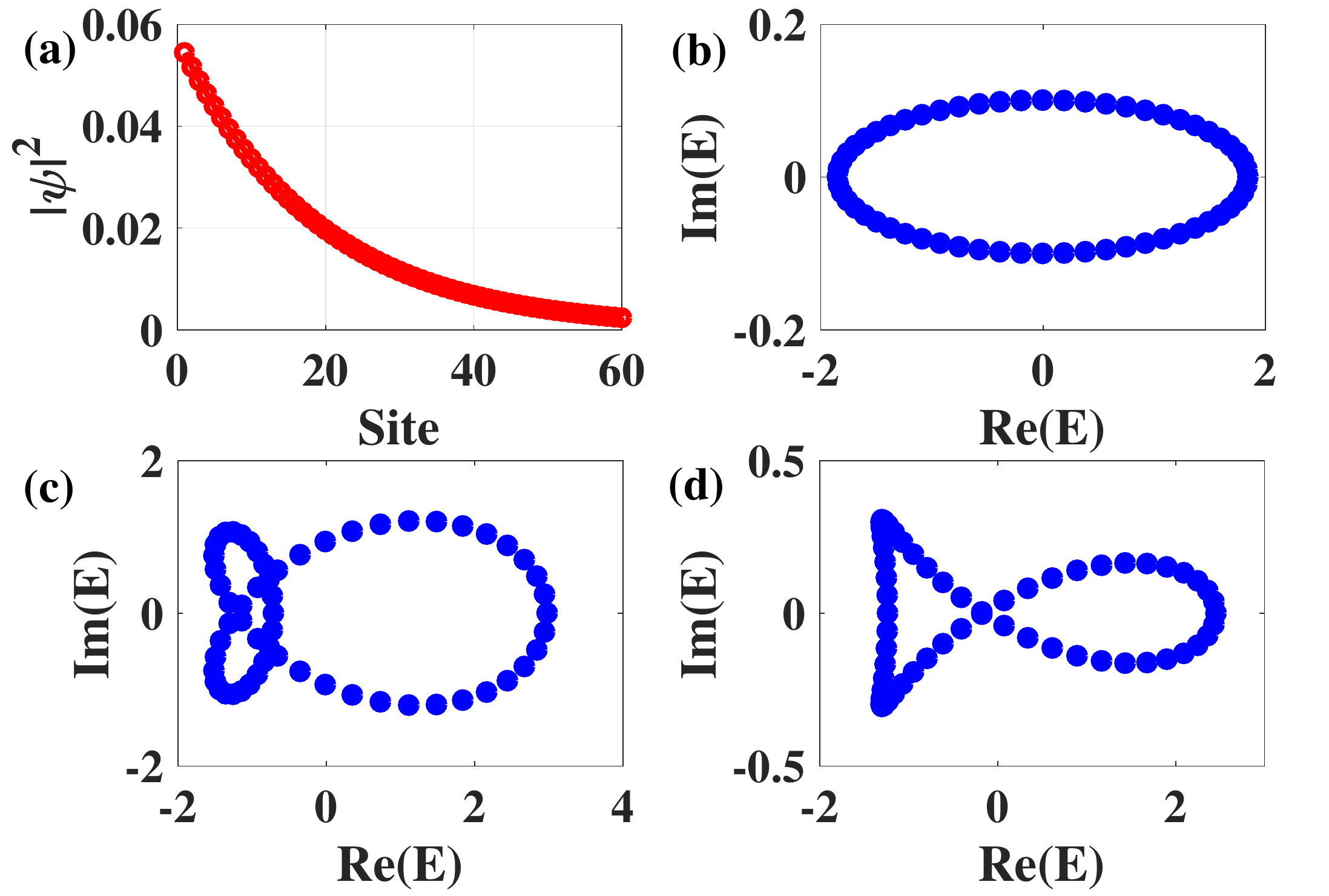}
\caption{(a) The profile of all eigenstates for different model in (b)-(d); (b) Energy spectra for general model with $p=1, q=1$ and $t_{1L}=1, t_{1R}=0.85$; (c) Energy spectra for general model with $p=2, q=1$ and $t_{1L}=1, t_{1R}=0.85, t_{2L}=1.2$; (d) Energy spectra for general model with $p=2, q=2$ and $t_{1L}=1, t_{1R}=0.85, t_{2L}=0.2, t_{2R}=0.4$. The boundary hopping parameters are determined by the specific boundary condition Eq.({\ref{Gcdt}}). Common parameters: $N=60, \mu=0.2$.}%
\end{center}
\label{fig2}
\end{figure}

\section{Results and discussions}
From the expression of wavefunction given by Eq.(\ref{GFFis}), we see that the distribution of wavefunction is only relevant to the value of $\mu$, but is irrelevant to the values of $t_{jL}$ and $t_{jR}$ as long as the specific boundary condition is fulfilled. This means that systems with very different spectrum structures may have the same wavefunctions.
When $|\mu| \neq1$, Eq.(\ref{GFFis}) suggests that non-Hermitian skin effect occurs as the distributions of wavefunctions decay exponentially from the left or right boundary. To see it clearly, in Fig. 2(a) we plot the distributions of wavefunctions for various systems fulfilled the specific boundary condition with $\mu=0.2$.  While the NHSE is distinct for small size systems, it becomes less distinct for large size systems as the wavefunctions approach extended states due to $|z_1|=\sqrt[N]{\mu }\mapsto 1$ when $N \rightarrow \infty$ \cite{GuoCX}.
Fig.\ref{fig2}~~~(b)-(d) display the spectra for different systems with the same sizes $N=60$. Although they display quite different spectrum structures, their wavefunctions are identical as shown in  Fig.\ref{fig2}~~~(a).

Next we shall discuss some special cases and display how the boundary parameter $\mu$ affects the spectra and wavefunctions.

\subsection{Hatano-Nelson model under the specific boundary condition}
\begin{figure}[btp]
\begin{center}
\includegraphics[width=0.46\textwidth]{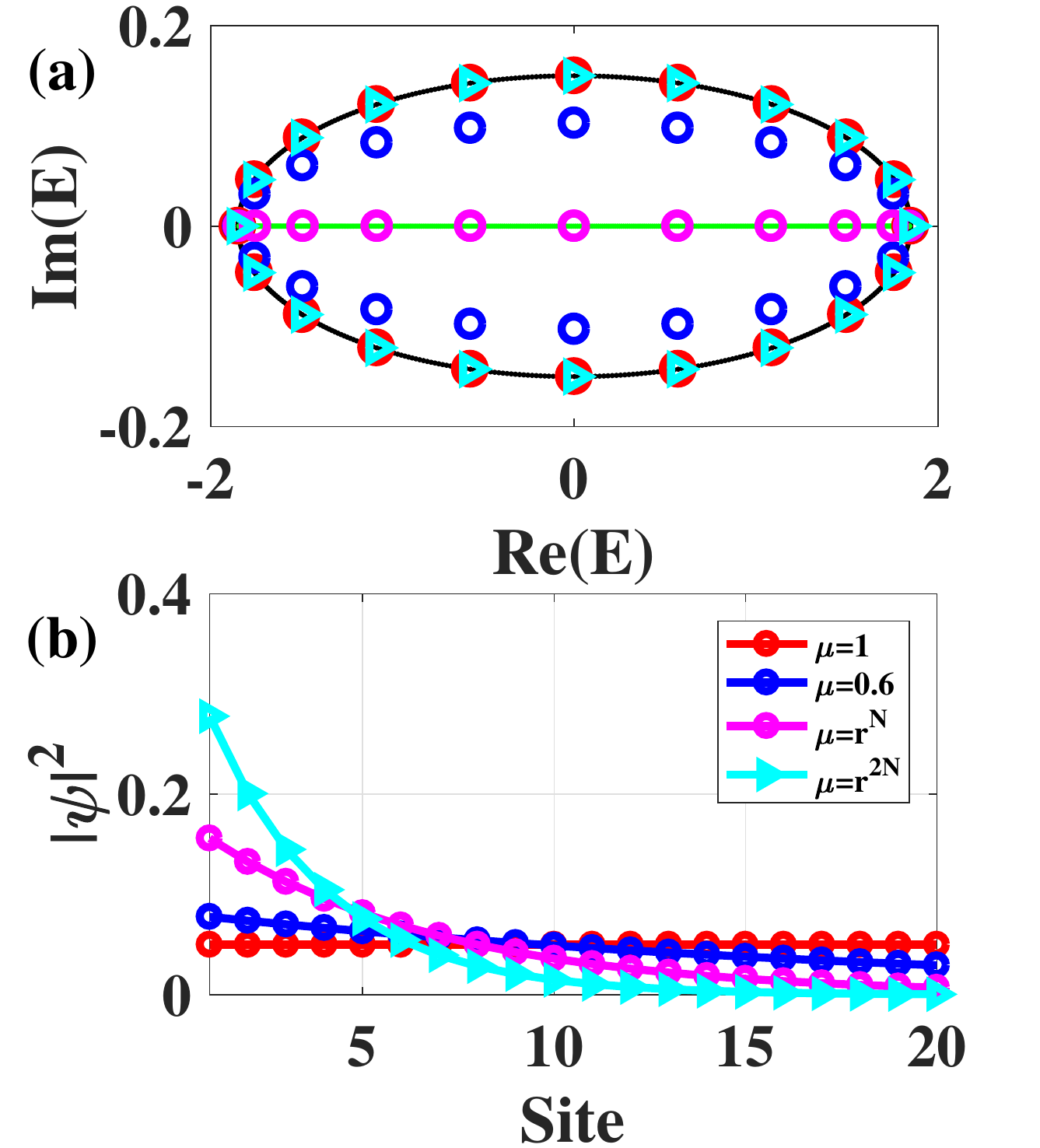}
\caption{Hatano-Nelson model under specific boundary conditions (a) Energy spectra with $\mu=1, 0.6, r^{N}, r^{2N}$ described by red circles, blue circles, magenta circles, cyan circles, respectively. The green and black line represents energy spectrum corresponding to  OBC and PBC case in the thermodynamic limit, respectively; (b) The profile of all eigenstates with $\mu=1, 0.6, r^{N}, r^{2N}$.  Common parameters: $t_{1L}=1, t_{1R}=0.85, N=20, r=\sqrt{t_{1R}/t_{1L}}$.}%
\end{center}\label{fig3}
\end{figure}
When $p=1$ and $q=1$, the general model reduces to the Hatano-Nelson model \cite{HatanoPRB,Hatano} under the specific boundary condition ${t_{1R}}/{\delta _{1R}}={\delta _{1L}}/{t_{1L}}=\mu$.
The corresponding eigenvalues from Eq.(\ref{GEZs}) with $p=1, q=1$ can be rewritten as
\begin{equation}
E=(t_{1L}\sqrt[N]{\mu }+\frac{t_{1R}}{\sqrt[N]{\mu }})\cos (\theta
)+i(t_{1L}\sqrt[N]{\mu }-\frac{t_{1R}}{\sqrt[N]{\mu }})\sin (\theta )
\end{equation}%
with $\theta=\frac{2m\pi}{N}$, and the eigenstates \textcolor{red}{are} given by Eq.(\ref{GFFis}).

There are some special situations in the specific boundary conditions as follows:

When $\mu=1$ (PBC), the eigenvalues can be expressed as
\begin{equation}
E=(t_{1L}+t_{1R})\cos (\theta)+i(t_{1L}-t_{1R})\sin (\theta )
\end{equation}
with $\theta=\frac{2m\pi}{N}$.

When $\mu=(\frac{t_{1R}}{t_{1L}})^{N/2}$, the eigenvalues are given by
\begin{equation}
E=2\sqrt{t_{1R}t_{1L}}\cos (\theta)
\end{equation}
with $\theta=\frac{2m\pi}{N}$. This special case is the so called modified PBC studied in Ref.\cite{Imura}.
We note that the spectra are similar to those under OBC, i.e.,
\begin{equation}
E=2\sqrt{t_{1R}t_{1L}}\cos (\theta)
\end{equation}
with $\theta=\frac{m\pi}{N+1}$. The corresponding wave functions under the  modified PBC exhibit NHSE.


When $\mu=(\frac{t_{1R}}{t_{1L}})^{N}$, 
the eigenvalues are given by
\begin{equation}
E=(t_{1L}+t_{1R})\cos (\theta)-i(t_{1L}-t_{1R})\sin (\theta ).
\end{equation}
Since the values $\theta$ appear always in pairs of $(\theta,-\theta)$ except the case of $\theta =
0,\pi$ $(\sin[0] = \sin[\pi] = 0)$, we find that the spectrum are the same as the spectrum under PBC, whereas the corresponding wave functions exhibit NHSE.
This special case is the so called pseudo-PBC studied in Ref.\cite{GuoCX}. This result is a little counterintuitive since we can get a Bloch-like spectrum even the translation invariance is broken by the boundary term. A straightforward interpretation is that the Hamiltonian under the pseudo-PBC $\hat{H}_{pPBC}$ can be transformed to a Hamiltonian $\hat{\tilde {H}}$ by carrying out a similar transformation, i.e., $S \hat{H}_{pPBC} S^{-1} = \hat{\tilde {H}}$, where $\hat{\tilde {H}}$ is identical to the original Hatano-Nelson model under PBC with $t_{1L}$ and $t_{1R}$ exchanged each other.

In Fig.3, we plot the energy spectra and the profile of eigenfunction with different $\mu$ for a fixed $N$. As shown in Fig.3(a), the energy spectra under pPBC ($\mu=r^N$) are the same as those under PBC ($\mu=1$), and both are located at energy spectra under PBC in the thermodynamic limit. In addition, the energy spectra under mPBC are located at spectra under OBC in the thermodynamic limit, which is consistent with our prediction. For $\mu>1$, the wavefunctions are localized on the left boundary, and the NHSE becomes more obvious as $\mu$ increase as displayed in Fig.3(b).

\subsection{Model with next-nearest-neighbor hopping}
Now we consider the model with $p=2$ and $q=2$ under the specific boundary condition ${t_{2R}}/{\delta _{2R}}={t_{1R}}/{\delta _{1R}}={\delta _{1L}}/{t_{1L}}={\delta _{2L}}/{t_{2L}}=\mu$. The corresponding eigenvalues from Eq.(\ref{GEZs}) with $p=2$ and $q=2$ can be rewritten as
\begin{equation}
\begin{split}
E=&\left[t_{1L}\sqrt[N]{\mu }+\frac{t_{1R}}{\sqrt[N]{\mu }}\right]\cos (\theta)\\
+ &\left[t_{2L}(\sqrt[N]{\mu })^2+\frac{t_{2R}}{(\sqrt[N]{\mu })^2}\right]\cos (2\theta) \\
+&i\left[ t_{1L}\sqrt[N]{\mu }-\frac{t_{1R}}{\sqrt[N]{\mu }}\right]\sin (\theta)  \\
+&i \left[(t_{2L}(\sqrt[N]{\mu })^2-\frac{t_{2R}}{(\sqrt[N]{\mu })^2}\right]\sin (2\theta)
\end{split}
\end{equation}%
with $\theta=\frac{2m\pi}{N}$, and the eigenstates are also given by Eq.(\ref{GFFis}).

When we apply PBC, i.e. $\mu=1$, the eigenvalues can be expressed as
\begin{equation}
\begin{split}
E=&(t_{2L}+t_{2R})\cos (2\theta)+(t_{1L}+t_{1R})\cos (\theta)\\
+&i\bigg[(t_{2L}-t_{2R})\sin (2\theta)+(t_{1L}-t_{1R})\sin (\theta)\bigg].
\end{split}
\end{equation}
Similar to the Hatano-Nelson model, we can also find the existence of a pseudo-PBC case for $\sqrt{\frac{t_{1R}}{t_{1L}}}=\sqrt[4]{\frac{t_{2R}}{t_{2L}}}=r$.
When $\mu=r^{2N}$, the eigenvalues can be expressed as
\begin{equation}
\begin{split}
E=&(t_{2L}+t_{2R})\cos (2\theta)+(t_{1L}+t_{1R})\cos (\theta)\\
-&i\bigg[(t_{2L}-t_{2R})\sin (2\theta)+(t_{1L}-t_{1R})\sin (\theta)\bigg]
\end{split}
\end{equation}
which are the same as those under PBC, while the corresponding wave functions exhibit NHSE.
Therefore, this special boundary condition is also named as the pseudo-PBC.

In Fig.4, we plot the energy spectra and the profile of eigenfunction with different $\mu$ for a fixed $N$. We can see that the energy spectra under pPBC are the same as those under PBC, while the wavefunctions exhibit NHSE obviously, which is consistent with our prediction.
\begin{figure}[tbp]
\begin{center}
\includegraphics[width=0.5\textwidth]{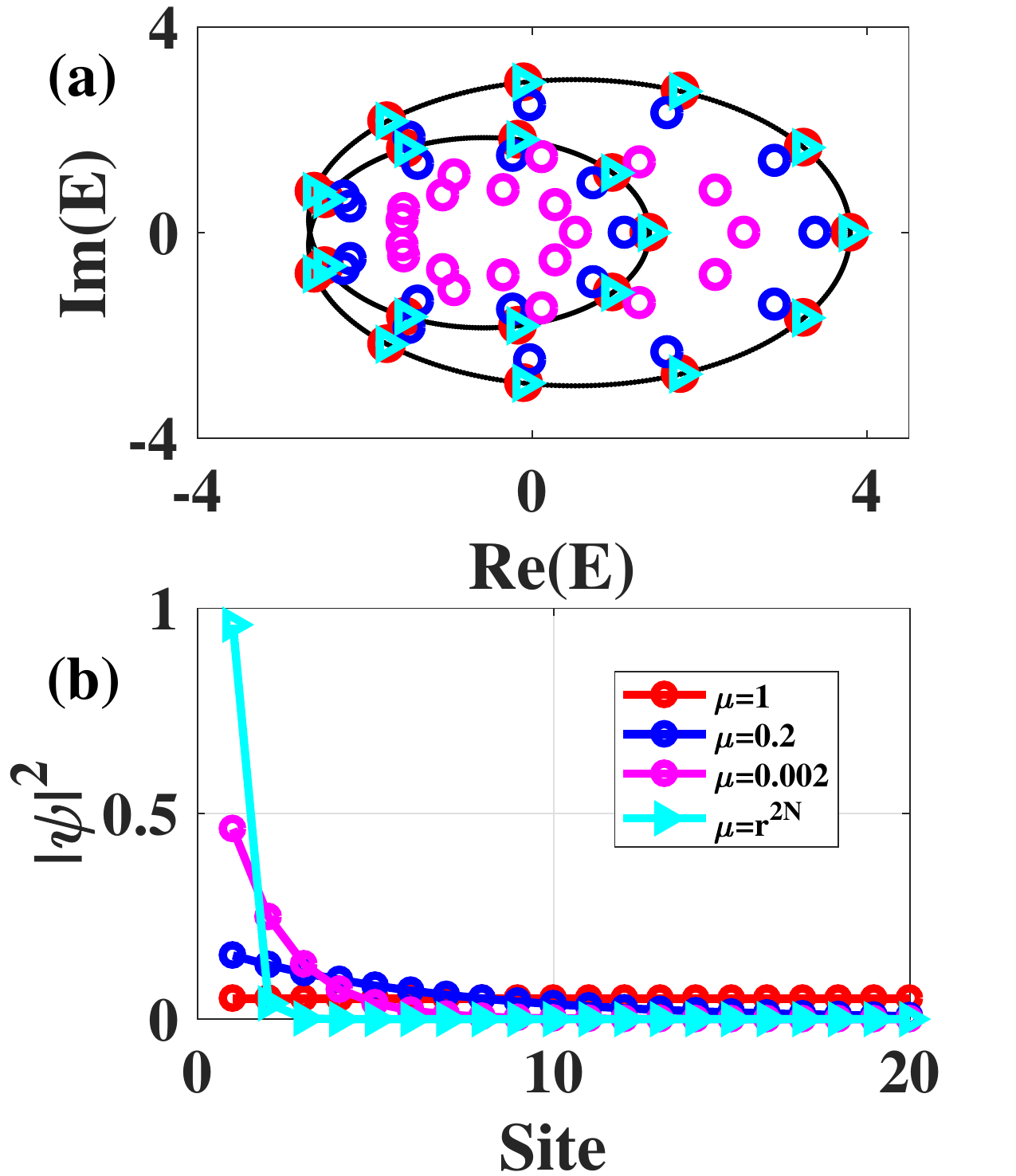}
\caption{General model with $p=2, q=2$ under specific boundary conditions (a) Energy spectra with $\mu=1, 0.2, 0.002, r^{2N}$ described by red circles, blue circles, magenta circles, cyan circles, respectively. The black line represents energy spectrum corresponding to  PBC case in the thermodynamic limit; (b) The profile of all eigenstates with $\mu=1, 0.2, 0.002, r^{2N}$.  Common parameters: $t_{1L}=1, t_{1R}=0.2, t_{2L}=2.5, t_{2R}=0.1, N=20, r=\sqrt{t_{1R}/t_{1L}}$.}%
\end{center}
\label{fig4}
\end{figure}

\subsection{General model with $p=q$}
For the general model with $p=q$ under the specific boundary conditions ${t_{qR}}/{\delta _{qR}}=\cdots={t_{1R}}/{\delta _{1R}}={\delta _{1L}}/{t_{1L}}=\cdots={\delta _{pL}}/{t_{pL}}=\mu$. the corresponding eigenvalues from Eq.(\ref{GEZs}) with $p=q$ can be rewritten as
\begin{equation}
\begin{split}
E=&\sum_{j=1}^{p}\bigg[(t_{jL}(\sqrt[N]{\mu })^j+\frac{t_{jR}}{(\sqrt[N]{\mu })^j})\cos (j\theta)\bigg]\\
+&i\sum_{j=1}^{p}\bigg[(t_{jL}(\sqrt[N]{\mu })^j-\frac{t_{jR}}{(\sqrt[N]{\mu })^j})\sin (j\theta)\bigg]
\end{split}
\end{equation}%
with $\theta=\frac{2m\pi}{N}$, and the eigenstates are given by Eq.(\ref{GFFis}).

When we apply PBC, i.e. $\mu=1$, the eigenvalues can be expressed as
\begin{equation}
\begin{split}
E=\sum_{j=1}^{p}\left[(t_{jL}+t_{jR})\cos (j\theta)\right]
+i\sum_{j=1}^{p}\left[(t_{jL}-t_{jR})\sin (j\theta)\right] .
\end{split}
\end{equation}
Similar to Hatano-Nelson model, we find the existence of a pseudo-PBC case as long as the following constrained relations
\begin{equation}
\sqrt[2j]{{t_{jR}}/{t_{jL}}}=r, ~~~ j=1,\cdots,p
\end{equation}
are fulfilled.
When $\mu=r^{2N}$, the eigenvalues can be expressed as
\begin{equation}
\begin{split}
E=\sum_{j=1}^{p}\left[(t_{jL}+t_{jR})\cos (j\theta)\right]
-i\sum_{j=1}^{p}\left[(t_{jL}-t_{jR})\sin (j\theta)\right]
\end{split}\nonumber
\end{equation}
which are the same as those under PBC, while the corresponding wave functions exhibit NHSE.
We also call this special case as the pseudo-PBC.

In Table 1, we display the energy spectra and eigenfunctions for non-Hermitian chains with asymmetric long-range hopping under different boundary conditions. It is noticed that pseudo-PBC exists only for the case of $p=q$, and we have $\theta=\frac{2m\pi}{N}$ $(m=1,\cdots,N)$ in the table.

\section{Conclusion}
In summary, we present exact solutions for general nonreciprocal chains with long-distance hopping under specific boundary conditions. Our analytical results indicate the existence of size-dependent NHSE. While the NHSE is distinct for small size system, it becomes less discernable in the large size limit. The wave functions are independent of hopping range, whereas the eigenvalue spectra are model dependent and display rich structures. We also find the existence of a special point called pseudo-PBC, for which the spectra are identical to periodic spectra when the hopping parameters meet certain conditions, while eigenstates display NHSE. Our exact solutions provide examples that the boundary terms
can dramatically change the bulk properties of non-Hermitian systems.
While both asymmetric and long-range hopping are hard to be realized in conventional quantum systems, electric circuits provide a platform to simulate non-reciprocal non-Hermitian systems \cite{Helbig,Liu}, which may be used to verify generalized bulk boundary correspondence and non-Hermitian skin effect. We expect that more interesting solutions can be found and be simulated in electric circuits in future works.


\onecolumngrid
\begin{center}
{\footnotesize{\bf Table 1.} Energy spectra and eigenfunctions for non-Hermitian chains with asymmetric long-range hopping under different boundary conditions. \\
\vspace{2mm}
\begin{tabular}{|c|c|c|c|}
  \hline
  Boundary condition &  Energy spectra & Eigenfunctions\\
  \hline
  Generalized boundary condition  & $E=\sum\limits_{j=1}^{p}  t_{jL}z_i^{j} +\sum\limits_{j=1}^{q}t_{jR}z_i^{-j}$ & $\Psi= \left(\psi_{1},\psi_{2},\cdots,\psi_{N} \right)^{T}$\\

  $(\delta _{1R},\cdots,\delta _{qR},\delta _{1L},\cdots,\delta _{pL})$  & (No general solution for $z_i$) & $(\psi_{n}=\sum_{i=1}^{q+p}(c_{i}z_{i}^{n}))$ \\

  \hline
  Specific boundary condition  & $E=\sum\limits_{j=1}^{p}  t_{jL}(\sqrt[N]{\mu}e^{i\theta})^{j}$
  & $\Psi= \left(\sqrt[N]{\mu }e^{i\theta},\cdots,\left( \sqrt[N]{\mu }e^{i\theta} \right)^{N} \right)^{T}$\\
  $(\frac{t_{1R}}{\delta _{1R}}\cdots=\frac{t_{qR}}{\delta _{qR}}=\frac{\delta _{1L}}{t_{1L}}=\cdots=\frac{\delta _{pL}}{t_{pL}}=\mu)$ & $+\sum\limits_{j=1}^{q}t_{jR}(\sqrt[N]{\mu}e^{i\theta})^{-j}$ & \\

  \hline
  {Periodic boundary condition} &   $E=\sum\limits_{j=1}^{p}  t_{jL}e^{i\theta j}+\sum\limits_{j=1}^{q}t_{jR}e^{-i\theta j}$  & $\Psi= \left(e^{i\theta},e^{i2\theta},\cdots, e^{iN\theta}  \right)^{T}$\\
  $(\mu=1)$ &  & (No non-Hermitian skin effect)\\

  \hline
  Pseudo-periodic boundary condition& $E=\sum\limits_{j=1}^{p}\left[(t_{jL}+t_{jR})\cos (j\theta)\right]$ & $\Psi= \left(r^2e^{i\theta},r^4e^{i2\theta},\cdots,\left( r^2e^{i\theta} \right)^{N} \right)^{T}$\\
for $p=q$ case$(\mu=r^{2N}$, $\sqrt[2j]{{t_{jR}}/{t_{jL}}}=r)$ & $-i\sum\limits_{j=1}^{p}\left[(t_{jL}-t_{jR})\sin (j\theta)\right]$ & (Exhibit non-Hermitian skin effect) \\
 \hline
\end{tabular}}
\end{center}

\twocolumngrid

\begin{acknowledgments}
The work is supported by National Key
Research and Development Program of China (2016YFA0300600), NSFC under Grants No.11974413, and the Strategic Priority Research Program of Chinese Academy of Sciences under Grant No. XDB33000000.
\end{acknowledgments}

\end{document}